\documentclass[letterpaper, 10 pt, conference]{ieeeconf}  

\IEEEoverridecommandlockouts                              

\usepackage{graphicx} 
\usepackage{amsmath} 
\usepackage{amssymb}  
\usepackage{siunitx}

\usepackage{lipsum}
\usepackage{caption}
\usepackage{subcaption}
\usepackage[acronym]{glossaries}
\usepackage{xcolor}
\usepackage{booktabs,tabularx}
\newcolumntype{L}[1]{>{\raggedright\arraybackslash}p{#1}}
\newcolumntype{I}[1]{>{\hangindent=4em}p{#1}}

\makeglossaries

\newacronym{MAPF}{MAPF}{Multi-Agent Path Finding}
\newacronym{ADG}{ADG}{Action Dependency Graph}
\newacronym{RPP}{RPP}{Replanning Prediction Problem}
\newacronym{SOC}{SOC}{Sum of Costs}

\title{\LARGE \bf
Should I Replan?\\Learning to Spot the Right Time in Robust MAPF Execution
}

\author{David Zahrádka$^{1,2}$, David Woller$^{1}$, Denisa Mužíková$^{1,2}$, Miroslav Kulich$^{1}$ and Libor Přeučil$^{1}$
\thanks{*This work was supported by European Union under the project Robotics and Advanced Industrial Production (reg. no. CZ.02.01.01/00/22\_008/0004590).
}%
\thanks{$^{1}$Czech Institute of Informatics, Robotics and Cybernetics, Czech Technical University in Prague, Czechia.
{\tt\small david.zahradka@cvut.cz}}%
\thanks{$^{2}$Faculty of Electrical Engineering, Czech Technical University in Prague, Czechia.}
}

\begin{document}

\nocite{IEEEexample:BSTcontrol}

\maketitle
\thispagestyle{empty}
\pagestyle{empty}

\begin{abstract}

During the execution of Multi-Agent Path Finding~(MAPF) plans in real-life applications, the MAPF assumption that the fleet's movement is perfectly synchronized does not apply.
Since one or more of the agents may become delayed due to internal or external factors, it is often necessary to use a robust execution method to avoid collisions caused by desynchronization.
Robust execution methods -- such as the Action Dependency Graph~(ADG) -- synchronize the execution of risky actions, but often at the expense of increased plan execution cost, because it may require some agents to wait for the delayed agents.
In such cases, the execution's cost can be reduced while still preserving safety by finding a new plan either by rescheduling (reordering the agents at crossroads) or the more general replanning capable of finding new paths.
However, these operations may be costly, and the new plan may not even lead to lower execution cost than the original plan: for example, the two plans may be the exact same.
Therefore, we estimate the benefit that can be achieved by single replanning in scenarios with delayed agents given an immediate state of the execution with a fully connected feed-forward neural network.
The input to the neural network is a set of newly designed ADG-based features describing the robust execution's state and the impact of potential delays, and the output is an estimated benefit achievable by replanning.
We train and test the network on a new labeled dataset containing $12,000$ experiments, and we show that our proposed method is capable of reducing the impact of delays by up to $94.6\%$ of the achievable reduction.

\end{abstract}

\section{Introduction}

\begin{figure}[htb]
    \centering
    \includegraphics[width=0.5\linewidth]{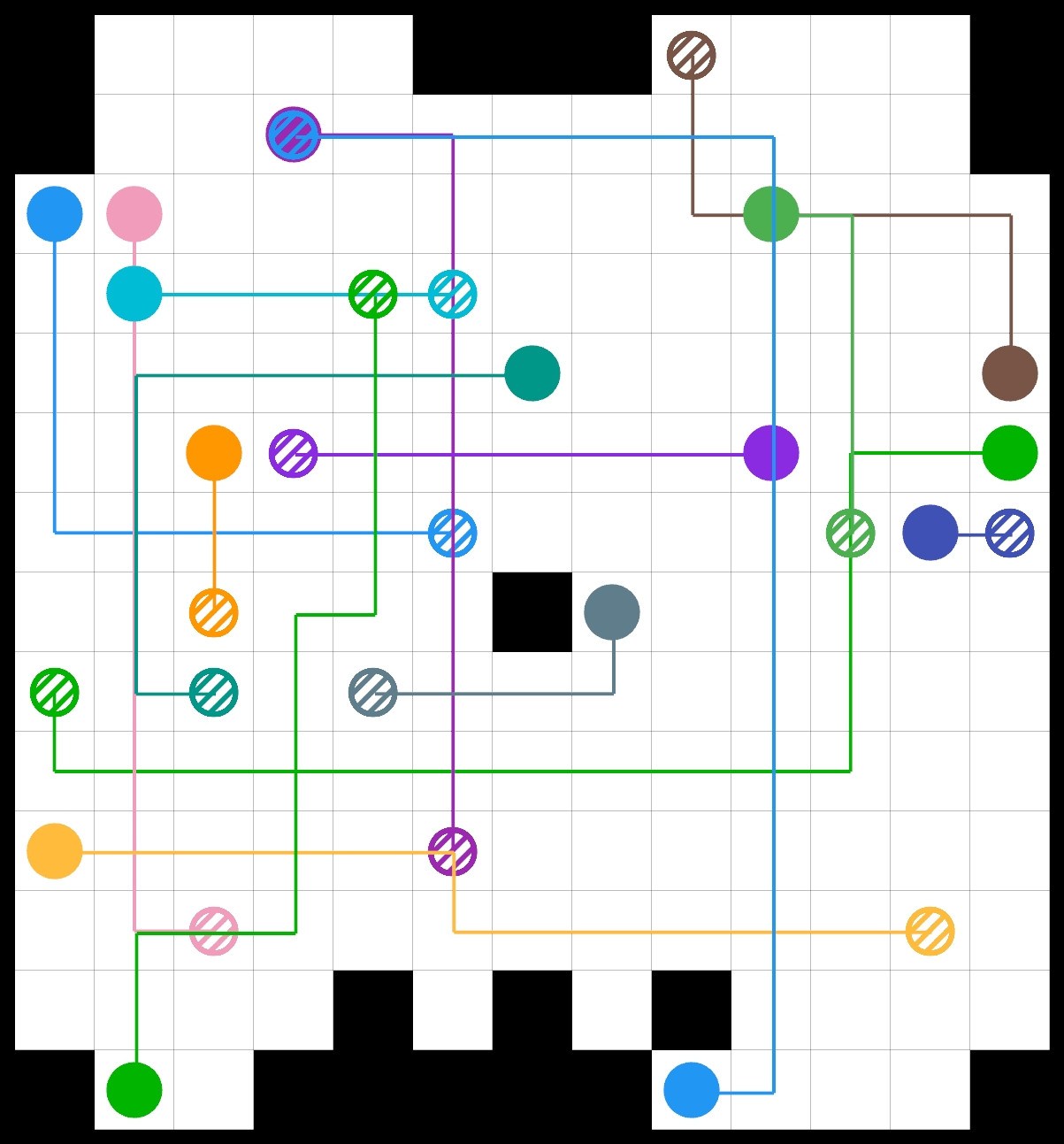}
    \caption{Example MAPF plan. Full circles are agents, hatched circles their goals, and lines represent paths.}
    \label{fig:example_solution}
\end{figure}

The goal of Multi-Agent Path Finding~(MAPF) is to find a set of collision-free paths in a shared environment with discrete space and time, represented as a graph, for a fleet of mobile agents such that the agents reach their given goals.
The problem is motivated by applications such as autonomous warehouses~\cite{wurman2008coordinating}, planning for swarms of Unmanned Aerial Vehicles~\cite{honig2018trajectory} and virtual network embedding~\cite{zheng2023improved}.
The goal is often to minimize some cost, such as the time when the last agent reaches its goal (\emph{makespan}) or the sum of lengths of all paths (\emph{Sum of Costs}).
While this \emph{optimization variant of MAPF} is an NP-hard problem for many criteria~\cite{surynek2010optimization}, recent solvers are able to solve instances with thousands of agents~\cite{okumura2024engineering}, albeit suboptimally.
An example of a MAPF plan can be seen in Fig.~\ref{fig:example_solution}.

The plans produced by MAPF solvers assume perfect synchronization of the agent fleet: all agents move at the same time, and execute exactly one action in each time step.
However, in practical applications, some agents may become unexpectedly delayed compared to the rest of the fleet, due to, for example, imprecise control, mechanical issues, or external factors such as a human entering the shared environment and blocking some robots.
Safe execution in the presence of unexpected delays can be achieved by using robust execution methods, such as MAPF-POST~\cite{honig2016multiagent} and Action Dependency Graph~(ADG)~\cite{Honig2019}.
These methods take a MAPF plan as an input, find temporal precedences between the actions of the agents and only allow executing an action when all of its precedences are complete, preventing both collisions and deadlocks.

However, with significant delays, it may happen that following the original plan leads to a longer execution than finding and executing a new plan, even if the original plan was optimal.
For example, if multiple agents are scheduled to move through a crossroad, but the first agent is delayed, the later-scheduled agents may also have to wait, increasing the delay's impact.
Changing the order of the agents on the crossroad (\emph{rescheduling}) could reduce the delay's impact.
\emph{Optimization of robust execution} was addressed by the development of robust execution methods capable of rescheduling, such as Switchable ADG~(SADG)~\cite{berndt2024receding} and Bidirectional Temporal Plan Graph~(BTPG)~\cite{su2024bidirectional}.
While SADG requires repeatedly solving a Mixed Integer Linear Problem to determine the agents' order, in BTPG, the ordering is defined by a first-come, first-served manner, although the resulting order may not be optimal.
A disadvantage of rescheduling itself is that it is not capable of producing alternative paths, which may be necessary to mitigate the impact of some delays. 
New paths can be found by \emph{replanning}, which is a more general approach that is also capable of rescheduling~\cite{kottinger2024introducing}.
However, it is also more costly.

A straightforward way to reduce execution duration (cost) is to reschedule/replan often, perhaps even every time step, provided that livelocks are prevented.
Since both rescheduling and replanning introduce some overhead, albeit small, frequent re-computation of the plan may cause it to accumulate over time.
Furthermore, the same plan may be re-created over and over again, because the solver may not be able to find a plan that would be better than the original one.
In some cases, the delay's effect may not even be mitigated at all.
Although the impact of rescheduling/replanning can be reduced by using a persistent execution method~\cite{Honig2019}, it is still more cost-effective if the frequency of rescheduling or replanning is not high. 
Predicting the potential benefit that rescheduling or replanning can bring in order to decide whether to find a new plan or not is, therefore, an important problem, which we investigate in this paper.

We use the framework proposed in~\cite{zahradka2025holistica} to monitor the progress of a robust execution.
We estimate the starting and completion times of all actions and measure the real values as the execution progresses.
The measured values are propagated through the ADG to improve the estimate of future actions.
For each action, we compute statistics, such as expected and real execution time, and the so-called slack, which represents how long does an agent have to wait for other agents or vice versa~\cite{zahradka2025holistica}.
We then use these statistics to train a neural network to predict the potential benefit of replanning, which we use to decide whether to replan or not.

The contributions of this paper are summarized as follows:
\begin{enumerate}
    \item we augment the ADG with novel metrics that capture the state of execution and the impact of potential delays in real time,
    \item we propose a data generation pipeline that produces a labeled dataset, enabling supervised training of a neural network to predict the benefit of replanning,
    \item we identify the most informative features and analyze the impact of a dynamic obstacle on execution cost,
    \item we experimentally demonstrate that the proposed approach recovers $94.6\%$ of the available cost savings.
\end{enumerate}

\section{Related Work}

\textbf{Robust execution methods} are used to safely execute MAPF plans in the presence of unexpected delays.
A straightforward approach called a Fully Synchronized Policy~\cite{ma2017multiagent} is to only assign actions belonging to the same time step, and only advancing to the next time step once all agents finished executing their action.
However, even independent agents have to wait for the rest of the fleet to complete their actions, increasing the duration of the execution.

More efficient methods include Minimal Communication Policy~\cite{ma2017multiagent}, RMTRACK~\cite{gregoire2018locallyoptimal}, MAPF-POST~\cite{honig2016multiagent}, ADG~\cite{Honig2019} and Kinodynamic Temporal Plan Graph~(TPG)~\cite{yan2025winktpga}.
A Minimal Communication Policy~\cite{ma2017multiagent} synchronizes the agents only on crossroads where the MAPF plan defines a strict order in which the agents move through.
RMTRACK~\cite{gregoire2018locallyoptimal} facilitates safe execution by ensuring that the trajectories remain in the same homotopy class in configuration space as the original planned trajectories.
MAPF-POST~\cite{honig2016multiagent} constructs a temporal network~(TN) with vertices representing events (agents entering locations) and edges representing precedence constraints between the events.
An event is scheduled only after all preceding events are scheduled to be completed.
A similar approach is used in ADG~\cite{Honig2019}, which is a TPG with vertices representing agents' actions and edges representing their precedences.
Safety is ensured by executing an action only after all preceding actions were completed.
Using actions instead of location-related events reduces the communication overhead between the central server and the agents.
The recent Kinodynamic TPG~\cite{yan2025winktpga} introduces kinodynamic constraints into ADG, which allows optimizing the velocities at which the agents move for smoother execution. 

Recent research also focuses on \textbf{rescheduling} the order in which agents enter crossroads to minimize the duration of the robust execution when an agent is delayed.
The problem was formulated as a Job Shop Scheduling Problem and solved with a metaheuristic in~\cite{zahradka2023solving}.
Another approach is to use any standard MAPF solver on a reduced planning graph built from the original paths of the agents to reschedule~\cite{kottinger2024introducing}. 

Rescheduling was also integrated into robust execution procedures themselves, allowing to minimize execution cost while maintaining deadlock and collision-free execution.
In~\cite{berndt2024receding}, SADG is introduced, which adds reverse precedences to each precedence between actions of different agents and formulates the problem of which of these two mutually exclusive edges to use as a Mixed Integer Linear Program.
In Switchable Edge Search~\cite{feng2024realtime}, an A*-style algorithm is used on SADG to solve the same problem.
Finally, BTPG is proposed in~\cite{su2024bidirectional}, which is built entirely before execution and schedules agents at crossroads in a first-come, first-served manner as the execution progresses.

\textbf{Machine learning}~(ML) was successfully applied to many different aspects of MAPF.
In~\cite{chen2024no}, multiple image-based ML models were used to select the best planning algorithm from a bank for a given instance. 
A popular application is to use ML inside existing solvers to improve their efficiency, such as selecting which part of the solution to destroy in Large Neighborhood Search~\cite{huang2022anytime}, or to repair the solution in~\cite{wang2025lns2+rl}.
Learning-based methods were also used as both decentralized solvers via reinforcement learning~\cite{damani2021primal2} and centralized solvers using neural networks~\cite{tang2025railgun}.
In this paper, we use ML to predict the potential benefit of replanning during the robust execution of a MAPF plan.

\section{Background \& Problem Formulation}

We start this section by recalling two fundamental concepts: \gls{MAPF} and \gls{ADG}, with notations adopted from~\cite{zahradka2025holistica}.
Then, we introduce the \gls{RPP}, which is the central focus of this paper.

\subsection{\glsentrylong{MAPF}}

\begin{figure}[h]
  \centering
  \begin{subfigure}{0.32\columnwidth}
    \centering
    \includegraphics[width=\linewidth]{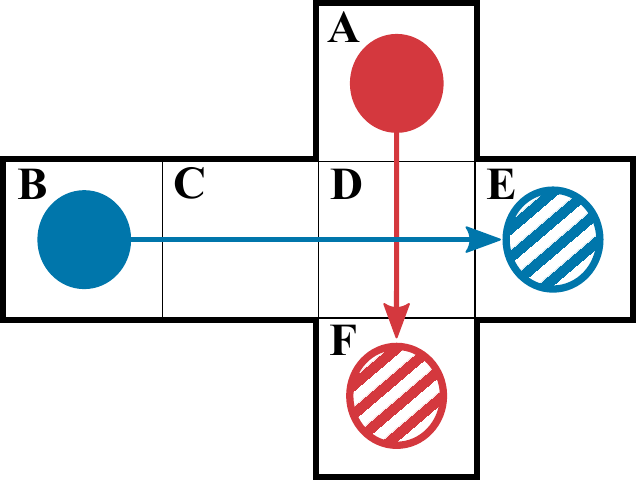}
    \caption{MAPF solution}
    \label{fig:MAPF_example}
  \end{subfigure}
  \hfill
  \begin{subfigure}{0.66\columnwidth}
    \centering
    \includegraphics[width=\linewidth]{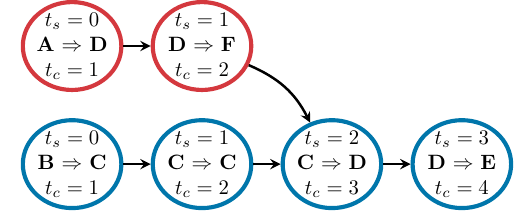}
    \caption{Corresponding ADG}
    \label{fig:ADG_example}
  \end{subfigure}
    \caption{Example MAPF instance with two agents}
  \label{fig:MAPF_ADG_examples}
\end{figure}

A \gls{MAPF} instance $M$ is defined as $M = (G, A)$, where $G=(V, E)$ is a graph, representing the shared environment, and $A$ is a set of agents operating on $G$. 
Each agent $k \in A$ has a starting location $s^k \in V$ and a goal location $g^k \in V$.
Then, a plan $\pi^k$ is a sequence of actions $a_i^k = (v_i, v_{i+1})$, where $a_i^k$ is the $i$-th action of agent $k$, representing movement from position $v_i \in V$ to $v_{i+1} \in V$ between discrete time steps $i$ and $i+1$.
If $v_i = v_{i + 1}$, agent is waiting in the same position.
Finally, the goal of \gls{MAPF} is to find a set of plans $\{\pi^k = (a_1^k,..,a_{|\pi^k|}^k)\mid k \in A\}$, so that $\forall k \in A: a_1^k = s^k, a_{|\pi_k|}^k = g^k$ and the objective function is minimized. 
We employ the commonly used criterion - \gls{SOC}, defined as
\begin{equation}
    SOC = \sum_{k \in A} |\pi^k|,
\end{equation}
which measures the total execution time across all agents.
An example MAPF instance is shown if Fig.~\ref{fig:MAPF_example}.

Each pair of plans must be conflict-free; we prohibit \textit{vertex}, \textit{swap}, \textit{following}, and \textit{cycle} conflicts \cite{stern2019multi}.
\gls{MAPF} solutions avoiding these conflicts are called 1-robust~\cite{Atzmon2020}, meaning that two different agents can't occupy the same vertex $v \in V$ at the same time step and in two consecutive time steps.

\subsection{\glsentrylong{ADG}}

Execution of MAPF plans in practice takes place in the continuous world instead of the discrete environment assumed in MAPF.
While it is possible to synchronize the agents so that they perform actions step-by-step, simulating discrete time, it prolongs the duration of the execution~\cite{ma2017multiagent}.
\gls{ADG}~\cite{Honig2019} facilitates robust execution of MAPF solutions with uncertain action durations or external factors by keeping track of prerequisities of all actions and allowing their execution only after the prerequisities are met. 

\gls{ADG} is a directed acyclic graph $G_{ADG} = (V_{ADG}, E_{ADG})$, where $V_{ADG}$ are all the actions $a_i^k$ in all plans $\pi_{k \in A}$ and $E_{ADG}$ are edges representing temporal dependencies between different actions.
There are two types of edges in $E_{ADG}$: type 1 edges $E_1$ for precedences between actions of a single agent and type 2 edges $E_2$ for dependencies between actions of different agents. 
An example ADG with $E_2 = \{(DF, CD)\}$ is shown in Fig.~\ref{fig:ADG_example}.
The red agent has to move from D to F before the blue agent can move from C to D in the next time step.

An extended version of the \gls{ADG}~\cite{zahradka2025holistica}, also keeps track of the estimated start time $\hat{t}_s(a_i^k)$ and estimated completion time $\hat{t}_c(a_i^k)$, where $a_i^k$ is the $i$-th action of agent $k$.
These values are periodically updated during the execution.
Once an action starts, the ADG records the actual start time $\bar{t}_s(a_i^k)$ and marks the action as running. 
After it finishes, ADG records its actual completion time $\bar{t}_c(a_i^k)$, sets $\hat{t}_c(a_i^k) = \bar{t}_c(a_i^k)$, marks the action as completed, and updates the estimated start and completion times of all subsequent actions.
Finally, let's denote the originally planned start and completion times as $t_s(a_i^k)$, $t_c(a_i^k)$, respectively.
At the beginning, the values are initialized as follows: $\forall k \in A, \forall a^k \in \pi^k:$
$\hat{t}_c(a^k) = t_c(a^k)$, $\bar{t}_c(a^k) = \infty$, $\hat{t}_s(a^k) = t_s(a^k)$, $\bar{t}_s(a^k) = \infty$.

\subsection{\glsentrylong{RPP}} 

We now define the problem of interest in this paper, the \gls{RPP}. 
It aims to predict whether replanning at a given time during the \gls{ADG}-controlled MAPF execution will be beneficial.
To enable this, various metrics within the \gls{ADG} are monitored and used as input features to the prediction.
Formally, let $x \in \mathbb{R}^d$ denote the feature vector. 
Let $y \in \mathbb{R}$ denote the predicted variable, representing the expected savings in \gls{SOC} that would be achieved by invoking replanning. 
The \gls{RPP} goal is to learn a mapping
\begin{equation} \label{eq:target}
    f : \mathbb{R}^d \to \mathbb{R}, \quad \text{such that } f(x) \approx y.    
\end{equation}
We formulate \gls{RPP} as a regression problem to allow for flexible thresholding and to preserve interpretability. 
A binary decision is then obtained by applying a threshold $\tau \in \mathbb{R}$ to the regression output: replan, if $f(x) \geq \tau$.
In our experimental setup, the regression target $y$ is defined as
\begin{equation}\label{eq:target_2}
    y = SOC^{ei} - SOC^{eir}_t,    
\end{equation}
where $SOC^{ei}$ denotes the \emph{executed} \gls{SOC} (in seconds) of the initial \gls{MAPF} plan  \emph{influenced} by external disturbances, and $SOC^{eir}_t$ denotes the \emph{executed} SOC of the same scenario if \emph{replanning} was triggered at time $t \in \mathbb{R}$. 
A positive value of $y$ therefore indicates that savings are expected.

\section{Method}

Now, we present the methodology for addressing the \gls{RPP}. 
Our main contribution is the definition of the features $x$ (~\ref{sec:features}). 
Sec.~\ref{sec:training} outlines the generation of the labeled dataset of observations $(x, y)$, and Sec.~\ref{sec:learning} describes the learning model used to approximate the regression function $f(x)$.

\subsection{Features definition}\label{sec:features}

Most proposed features are derived from the current \gls{ADG} state during \gls{MAPF} execution. 
The feature vector $x$ changes whenever the \gls{ADG} updates, ensuring that it reflects the latest execution state. 
Replanning may be triggered at each \gls{ADG} update based on the predicted benefit $f(x)$.

Table~\ref{tab:features} lists the 18 distinct features along with their formal definitions. 
Several features are parameterized by $n$, which specifies the number of past actions over which the feature is evaluated.
For clarity, the table is split into three groups: MAPF instance properties, MAPF solution properties, and current ADG state properties. 
The main idea is to detect present or expected deviations of \gls{ADG}-controlled execution at the current time $t$ from the original \gls{MAPF} plan. 

First, we introduce some auxiliary notations and symbols.
Let $p^k(t)$ be the \textit{progress} of agent $k \in A$, corresponding to the index of the last finished action at time $t$ in its plan $\pi^k = (a_1^k, .., a_{|\pi^k|}^k)$.
Then, the last $n$ finished actions of agent $k$ at time $t$ can be defined as $P(k, t, n) = \{a_{\max(1, p^k(t) - n + 1)}^k, .., a_{p^k(t)}^k\}$.
If no action has finished yet, $p^k(t) = 0$ and $P(k, t, n) = \emptyset$.
Analogously, let $e^k(t)$ be the index of the last action assigned to agent $k$.
If $e^k(t) > p^k(t)$, agent $k$ is currently executing the action $a^k_{e^k(t)}$. 
Otherwise, agent $k$ is either finished (if $p^k(t) = |\pi^k|$) or blocked by the \gls{ADG}, meaning that he is waiting for another agent.
Again, let's denote the last $n$ actions assigned to agent $k$ at time $t$ as $E(k, t, n) = \{a_{\max(1, e^k(t) - n + 1)}^k, .., a_{e^k(t)}^k\}$.
If no action was assigned yet, $e^k(t)=0$ and $E(k, t, n) = \emptyset$.

We define several features that capture \textit{action delay}, quantifying the excess execution time of the considered actions. 
This is computed by measuring the deviation between the planned duration of an action, $d(a) = t_c(a) - t_s(a)$, and its actual duration, $\bar{d}(a) = \bar{t}_c(a) - \bar{t}_s(a)$, or expected duration, $\hat{d}(a) = \hat{t}_c(a) - \hat{t}_s(a)$. 
In addition, we define \textit{plan delay} features, which evaluate the difference between the actual or expected completion times ($\bar{t}_c, \hat{t}_c$) and the planned completion times $t_c$. 
Unlike action delay features, plan delay features also capture indirect effects, such as excess waiting time for agents without an assigned action.

One of the features, \textit{highest slack increase}, is adopted from~\cite{zahradka2025holistica}. 
It is designed to identify when the current execution delay is likely to cause excess waiting in the future.
For every type 2 edge $e \in E^2$, $e = (a_i^{k}, a_j^{l})$ between actions of agents $k$ and $l$, we define \textit{planned slack} as $\delta(e)=t_c(a_i^{k}) - t_c(a_{j - 1}^{l})$ and \textit{expected slack} as $\hat{\delta}(e) = \hat{t}_c(a_i^{k}) - \hat{t}_c(a_{j - 1}^{l})$.
For agent $l$, a positive \textit{planned slack} value means that, according to the initial plan, agent $l$ will wait for agent $k$ for $\delta(e)$ seconds.
If the \textit{expected slack} increases relative to \textit{planned slack}, agent $l$ might need to wait longer than originally planned.

Let's also define the set of satisfied dependencies between two different agents in $E_{ADG}$ as $\bar{E}^2$.
These are edges, where action $a'$ is waiting for action $a$ to finish, which already happened: $\bar{E}^2 = \{e = (a, a'): e \in E^2 \wedge \bar{t}_c(a) = \hat{t}_c(a)\}$.
Finally, let $[\![ f ]\!] = 1$, if formula $f$ is true, 0 otherwise.

\begin{table}[h]
\centering
\caption{Features $x$, defined for the \gls{RPP}}
\label{tab:features}
\footnotesize
\setlength{\tabcolsep}{4pt}
\renewcommand{\arraystretch}{1.1}
\begin{tabularx}{\columnwidth}{@{}L{0.33\columnwidth} X@{}}
\toprule
\textbf{\textit{Name}}                   & \textbf{Description} \\
\midrule
\textit{map height, map width}           & Dimensions of the grid environment                \\
\textit{agents count}                     & No. of agents in the \gls{MAPF} instance: $|A|$   \\
\midrule
\textit{planned SOC}                     & $SOC = \underset{k \in A}{\sum} |\pi^k|$  \\
\textit{planned makespan}                & $\underset{k \in A}{\max} |\pi^k|$  \\
\midrule
\textit{replan time}                     & Current time $t$; potential replanning time \\
\textit{unfinished agents count}          & No. of agents not in their goals yet: $\underset{k \in A}{\sum} [\![ p^k(t) \neq |\pi^k| ]\!]$ \\
\textit{progress gap}                    & Max. progress difference between two agents: 
                                    $\underset{k \in A}{\max} (p^k(t))  - \underset{k \in A}{\min} (p^k(t))$\\
\textit{highest plan delay}              & Highest plan delay among last finished actions:
                                    $\underset{k \in A}{\max} (\bar{t}_c(a^k_{p^k(t)}) - t_c(a^k_{p^k(t)}))$  \\
\textit{highest exp. plan delay}         & Highest expected plan delay among currently assigned actions: 
                                    $\underset{k \in A}{\max} (\hat{t}_c(a^k_{e^k(t)}) - t_c(a^k_{e^k(t)}))$ \\ 
\textit{total plan delay}                & Total plan delay over all last finished actions:
                                    $\underset{k \in A}{\sum} (\bar{t}_c(a^k_{p^k(t)}) - t_c(a^k_{p^k(t)}))$  \\
\textit{total exp. plan delay}           & Total expected plan delay over all currently assigned actions:
                                    $\underset{k \in A}{\sum} (\hat{t}_c(a^k_{e^k(t)}) - t_c(a^k_{e^k(t)}))$ \\
\textit{highest action delay (n)}        & Highest total delay of last $n$ finished actions:
                                    $\underset{k \in A}{\max} (\underset{a \in P(k, t, n)}{\sum}(\bar{d}(a) - d(a)))$  \\
\textit{highest exp. action delay (n)}   & Highest expected total delay of last $n$ assigned actions:
                                    $\underset{k \in A}{\max} (\underset{a \in E(k, t, n)}{\sum}(\hat{d}(a) - d(a)))$  \\
\textit{total action delay (n)}          & Total delay of last $n$ finished actions:
                                    $\underset{k \in A}{\sum} (\underset{a \in P(k, t, n)}{\sum}(\bar{d}(a) - d(a)))$  \\
\textit{total exp. action delay (n)}     & Total expected delay of last $n$ assigned actions:
                                    $\underset{k \in A}{\sum} (\underset{a \in E(k, t, n)}{\sum}(\hat{d}(a) - d(a)))$  \\
\textit{highest slack increase}          & Highest slack increase over all unmet dependencies between agents: 
                                    $\underset{e \in E^2 \setminus \bar{E}^2}{\max} (\hat{\delta} (e) - \delta (e))$  \\
\textit{waiting agents count}             & Number of agents not currently executing an action:
                                    $\underset{k \in A}{\sum} [\![ e^k(t) = p^k(t)]\!]$ \\
\bottomrule
\end{tabularx}
\end{table}

\subsection{Labeled training and testing dataset}\label{sec:training}

In order to train our prediction model, we design a labeled dataset of observations $(x,y)$.
Problem instances are randomly generated with varying numbers of agents, map sizes, and multiple random start-goal assignments for the agents.
For each instance, we use multiple combinations of varying replanning times $t$ and dynamic obstacles to obtain more results.
The entries in the dataset then contain the measured \textit{execution} cost of an optimal 1-robust plan \textit{influenced} by the dynamic obstacle $SOC^{ei}$, where no replanning occurred, the execution cost of the same plan disturbed by the same obstacle, when \textit{replanning} occurred at time $t$: $SOC^{eir}_t$, and the feature vector $x$ as measured at the time $t$.
From these, the regression target $y$ is computed (Eq.~\ref{eq:target_2}).

Additionally, we measure the execution time without any disturbance: $SOC_e$ and with replanning, including the \emph{planning overhead} caused by the replanning process: 
\begin{equation}\label{eq:overhead}
    SOC^{eirp}_t = SOC^{eir}_t + 
    t_r \times 
    \sum_{k \in A} 
    \big[\![ p^k(t) \neq |\pi^k| ]\!\big],
\end{equation}
where $t_r$ is the runtime of the solver and the summation counts the agents that have not yet finished at time $t$. 
This term quantifies the additional SOC lost while replanning.
We do not use $SOC^{eirp}_t$ to define the regression target $y$ in Eq.~\ref{eq:target_2}, as we aim to keep the key experiments independent of the specific solver employed.
It is used only to provide further insight into the performance of our method.

\subsection{Learning model}\label{sec:learning} 

We employ a fully connected feed-forward neural network, implemented using scikit-learn~\cite{scikit-learn} and TensorFlow~\cite{tensorflow2015-whitepaper}, to learn the mapping $f : \mathbb{R}^d \rightarrow \mathbb{R}$. 
It consists of an input layer, three hidden layers with $64$, $32$, and $16$ neurons, respectively, each using ReLU activation, and a single linear output neuron predicting the expected SOC savings $f(x)$.
Prior to training, we apply the \texttt{RobustScaler} from scikit-learn to both input $x$ and $y$, centering their distributions at zero and scaling it so that the interquartile range equals one.
The network is trained to minimize the mean absolute error (MAE) loss:
\begin{equation}
    \mathcal{L}_{\mathrm{MAE}} = 
    \frac{1}{N} \sum_{i=1}^{N} \big| f(x_i) - y_i \big|,
\end{equation}
where $N$ is the number of training samples. 
Together, robust scaling and MAE loss reduce the model's sensitivity to outliers.
We use the Adam optimizer with the step decay learning rate $\eta(s)$ defined as
\begin{equation}
    \eta(s) = \eta_0 \, \gamma^{\lfloor s / s_{\mathrm{decay}} \rfloor},
\end{equation}
where $s$ is the current training step, $\eta_0$ the initial learning rate, $\gamma$ the decay rate, and $s_{\mathrm{decay}}$ the decay step size. 
Training is performed in mini-batches of size $B$ with early stopping based on validation to prevent overfitting. 
The final model was designed and tuned using $k$-fold cross-validation.

\section{Experimental Setup}

In this chapter, we describe the experimental environment (Sec. \ref{sec:environment}), detail the instance generation (Sec.~\ref{sec:instances}), and present the parameters related to the dataset handling, learning model design, and the tuned hyperparameters (Sec.~\ref{sec:learning_params}).

\subsection{Experimental environment} \label{sec:environment}
We use a 1-robust ECBS~\cite{barer2014suboptimal} solver with suboptimality bound $1.0$ to find both optimal initial plans and new optimal plans when replanning.
To measure their execution cost, we use a simulated environment~\cite{zahradka2025holistica}.
It consists of a central robust execution server that manages threads of all agents.
The server employs ADG to decide whether each action is safe to execute and also to monitor the execution by recording various execution metrics.
Each action takes exactly $\SI{1}{\second}$ and the agents execute them perfectly, with only a small overhead generated by the centralized execution architecture.
However, while AGD prevents inter-agent collisions, a dynamic obstacle (an intrusion) may appear in the environment and block a vertex that is currently free and no agent is moving in or out of it.
Agents avoid colliding with the dynamic obstacle by checking whether the goal location of an action is free when starting its execution.
If there is an obstacle, the agent starts moving only after it clears.
This way, delays are introduced to the otherwise perfectly moving agents.

The dynamic obstacle appears and disappears at random times in a random vertex.
We select a random appearance time at least $\SI{3}{\second}$ before the plan is estimated to be completed, as indicated by makespan.
It appears in a currently unoccupied vertex that is to be visited by an agent $\SI{3}{\second}$ after the appearance time.
The $\SI{3}{\second}$ are a time buffer that ensures that the dynamic obstacle has sufficient opportunities to enter the vertex.
The obstacle remains in the vertex before its randomized disappearance time elapses, which is no sooner than $\SI{3}{\second}$ after it appears to guarantee that it will affect the blocked agent, and no later than the plan makespan.
The appearance and disappearance times and the vertex are all controlled by the dynamic obstacle randomization seed.

Each measured execution cost is obtained through an independent simulator run.
To mitigate stochastic errors caused by multithreading, which may influence the results measured by the simulator, we repeat each experiment three times and select the minimal measured values.

\subsection{Instances}\label{sec:instances}

\begin{figure}
    \centering
    \begin{subfigure}[c]{0.49\columnwidth}
        \centering
        \includegraphics[height=3.0cm]{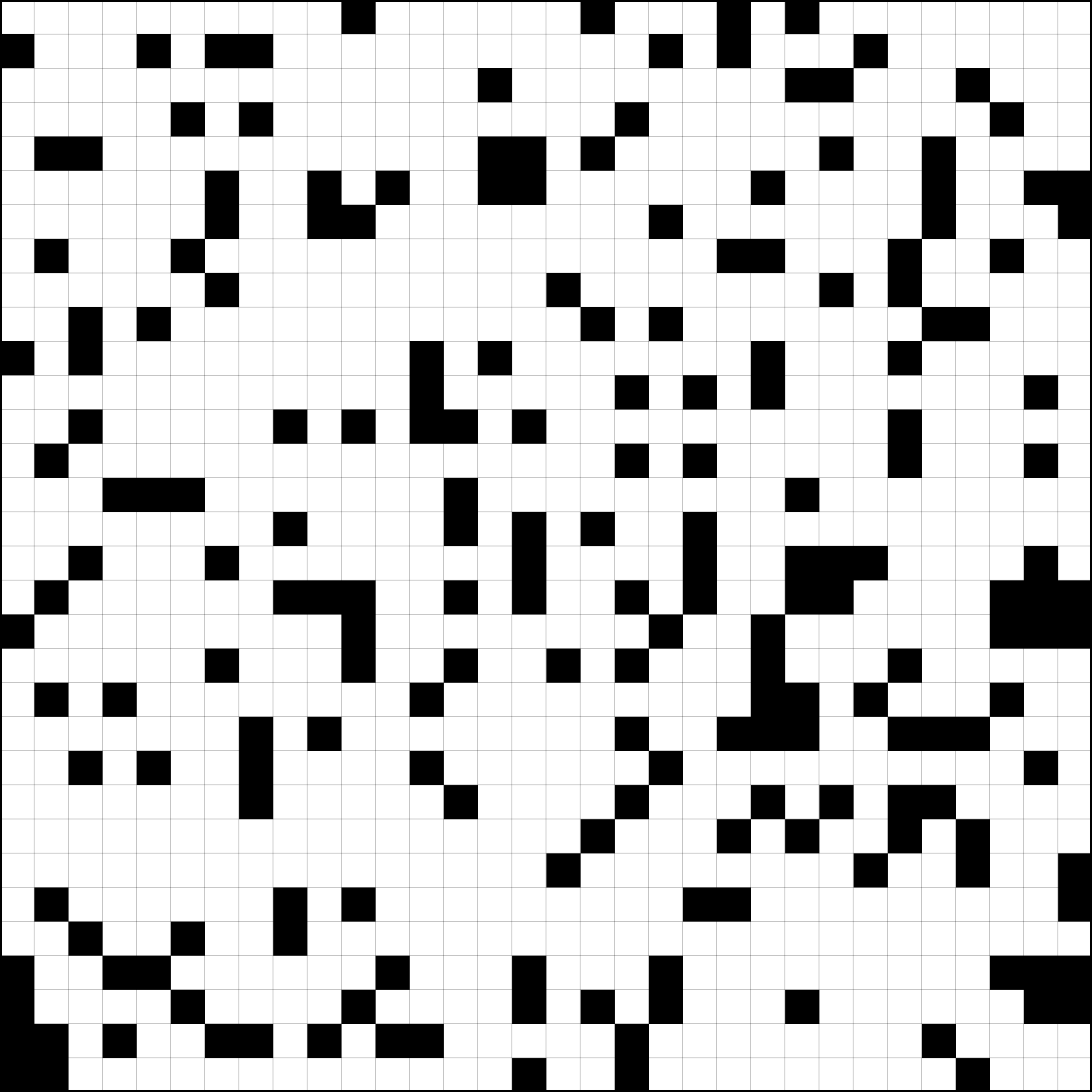}
        \caption{\texttt{random-32-32-20}}
    \end{subfigure}
    \begin{subfigure}[c]{0.49\columnwidth}
        \centering
        \includegraphics[height=3.0cm]{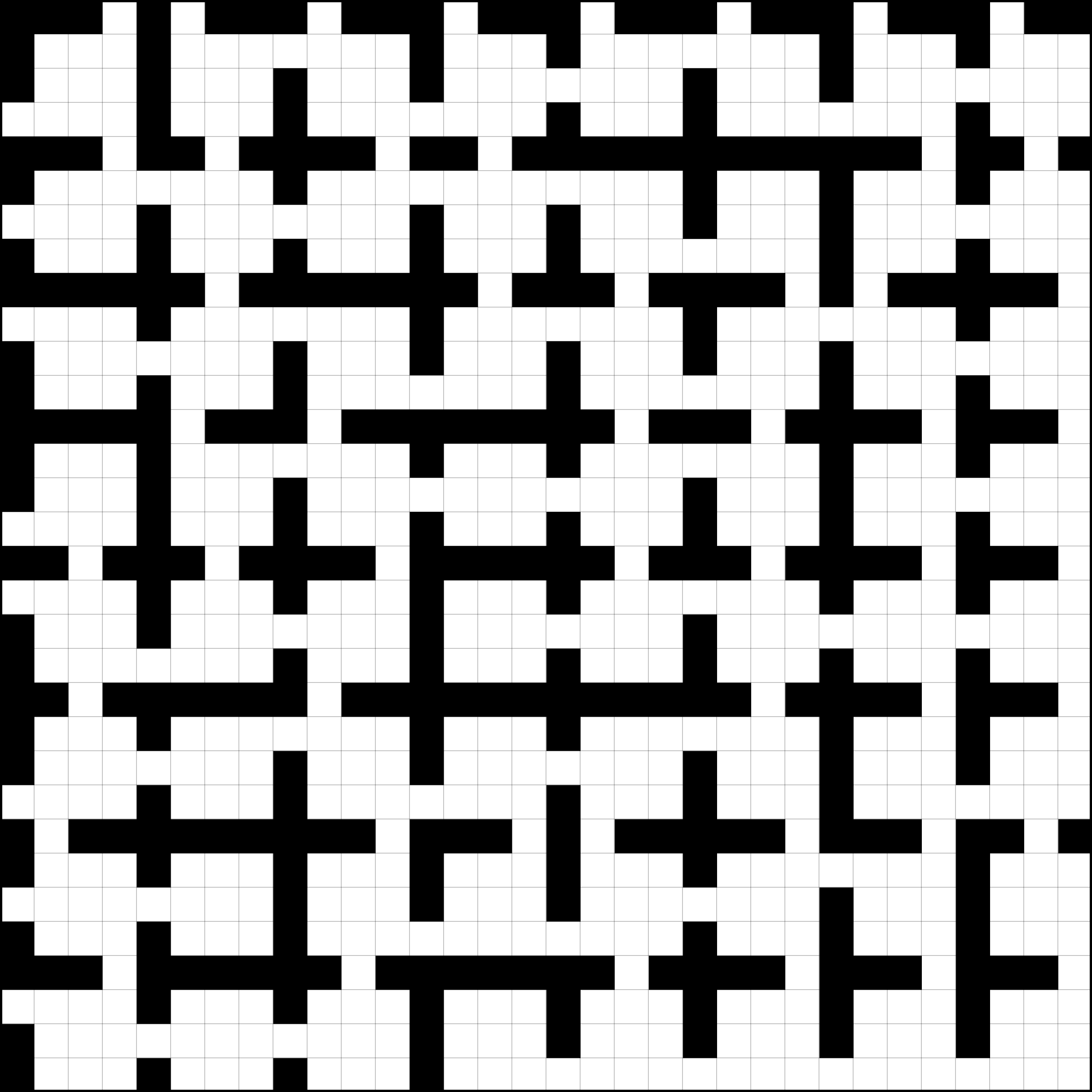}
        \caption{\texttt{room-32-32-4}}
    \end{subfigure}
    \begin{subfigure}[c]{0.49\columnwidth}
        \centering
        \includegraphics[height=3.0cm]{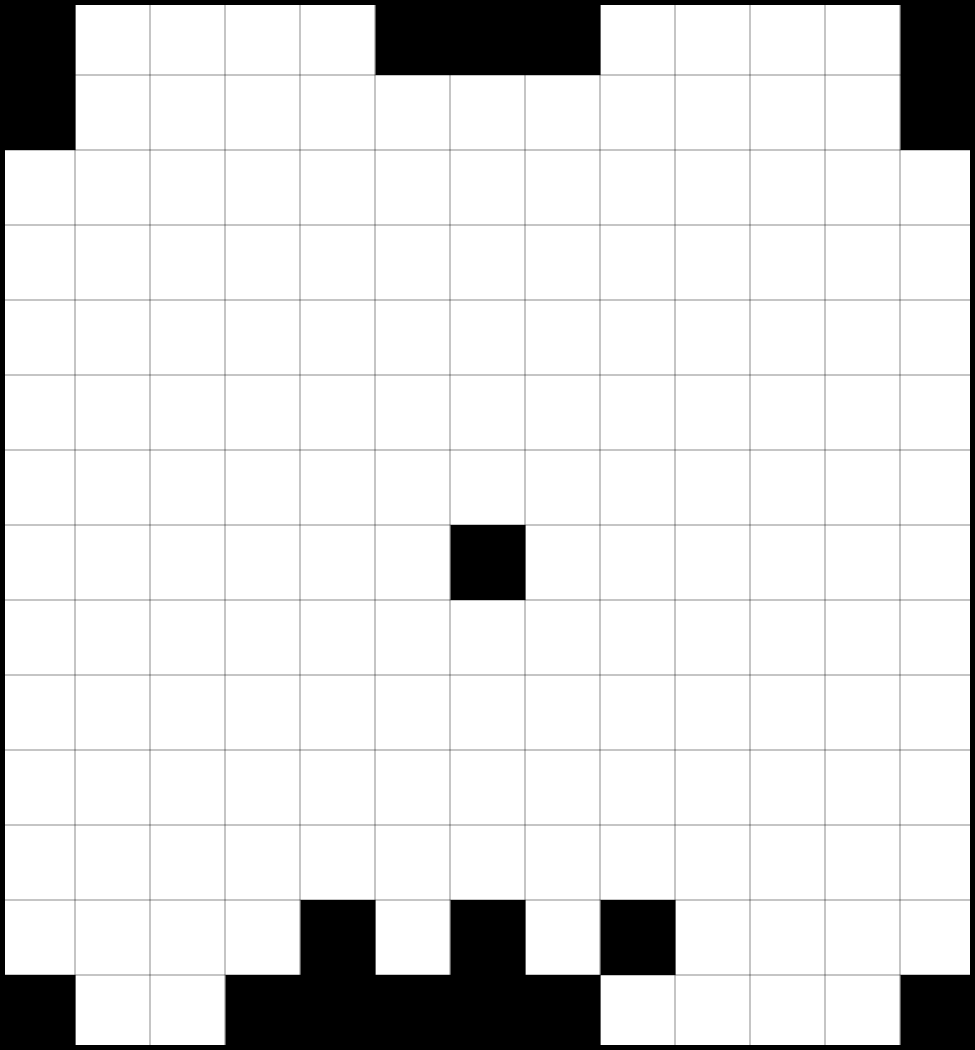}
        \caption{\texttt{lab}}
    \end{subfigure}
    \begin{subfigure}[c]{0.49\columnwidth}
        \centering
        \includegraphics[height=3.0cm]{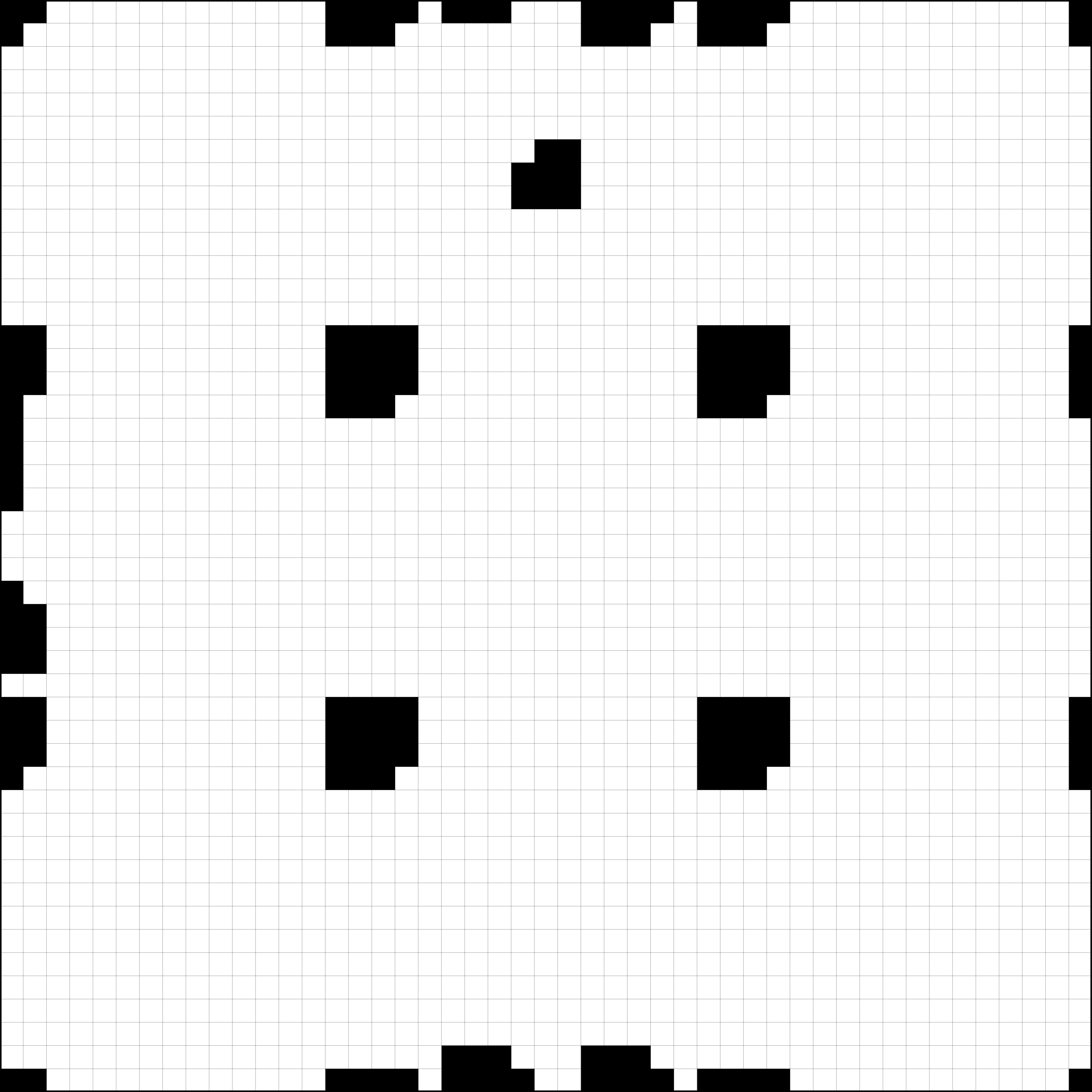}
        \caption{\texttt{arena}}
    \end{subfigure}
    \caption{Maps used in experiments}
    \label{fig:maps}
\end{figure}

We used four maps of various sizes: \texttt{random-32-32-20} and \texttt{room-32-32-4}, which are maps with $32\times32$ cells taken from the MovingAI dataset~\cite{stern2019multi} and \texttt{arena} and \texttt{lab} from~\cite{zahradka2025holistica} with dimensions $49\times49$ and $13\times14$, respectively.
The maps can be seen in Fig.~\ref{fig:maps}.
For each map, we used three different numbers of agents: $\{15, 20, 25\}$ for \texttt{arena} and $\{5, 10, 15\}$ for \texttt{lab}, \texttt{random} and \texttt{room} maps.
For each map and number of agents, we generated 20 instances that were solvable with the optimal 1-robust solver within $\SI{60}{\second}$.
We run each instance with $5$ dynamic obstacle randomization seeds and $10$ replanning seeds, producing $3000$ different experiments per map, thus $12000$ experiments in total.
Exactly one replanning time $t$ was always generated before the obstacle's appearance, and the rest after.
After generation, the dataset is split into training set $\mathcal{D}_{train}$ and test set $\mathcal{D}_{test}$, so that $|\mathcal{D}_{train}| = 8400$ and $|\mathcal{D}_{test}| = 3600$. 

\subsection{Learning parameters}\label{sec:learning_params}

We employed $5$-fold cross-validation on $\mathcal{D}_{train}$ for model design and hyperparameter tuning. 
The final model is trained with a batch size of $B=64$, meaning that weight updates are computed after processing 64 samples. 
Training is performed for at most $E_{\max}=500$ epochs, with early stopping triggered if the validation loss does not improve for $E_{\text{stop}}=100$ consecutive epochs. 
The validation loss is computed on a randomly sampled subset of the training set using the validation split ratio $\rho_{\mathrm{val}}=0.2$.
The remaining hyperparameters, that were already introduced in Sec.~\ref{sec:learning}, were set as follows: $\eta_0 = 0.001$, $\gamma = 0.96$, and $s_{\mathrm{decay}} = 100$.

\section{Results \& Discussion}

We evaluate the complete pipeline for the \gls{RPP} through a series of experiments. 
First, we compare different execution scenarios (Sec.~\ref{sec:scenarios}) to assess the influence of intrusion, random replanning and computational overhead. 
Next, we analyze the prediction performance of the proposed model (Sec.~\ref{sec:predicted_savings}). 
Then, we evaluate the realized savings when the model is used to trigger replanning (Sec.~\ref{sec:realized_savings}). 
Finally, we assess the contribution of individual features (Sec.~\ref{sec:features_importance}).

\subsection{Execution Scenarios}\label{sec:scenarios}

\begin{figure}[ht]
  \centering
  \includegraphics[width=\columnwidth]{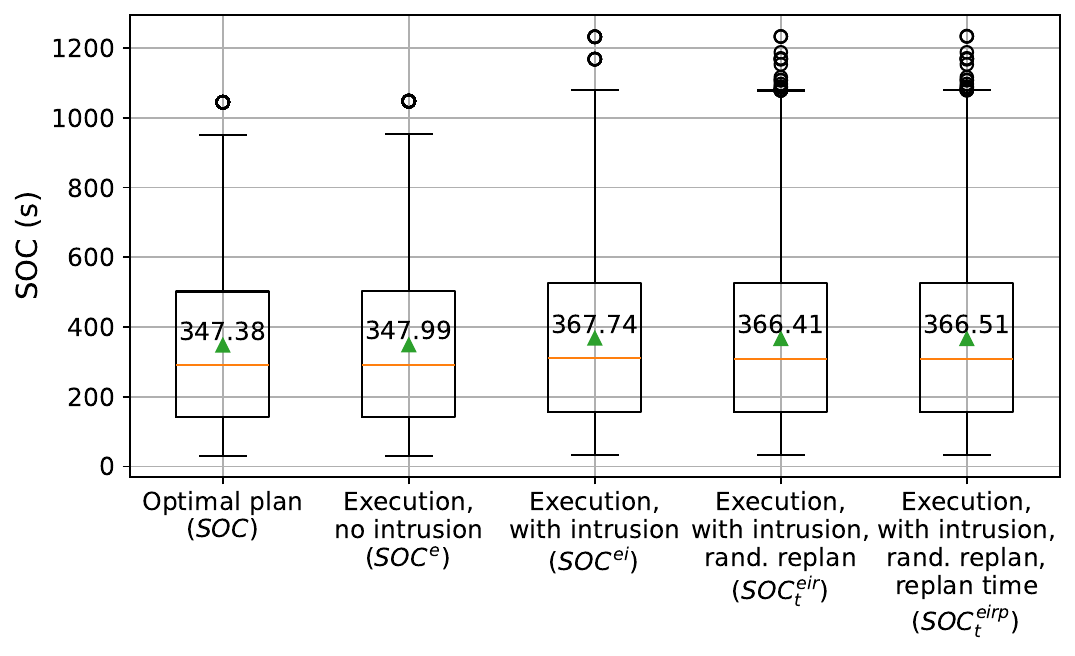}
  \caption{SOC across all scenarios}
  \label{fig:soc_setups}
\end{figure}

Three simulator scenarios are required to generate a single experiment, corresponding to one labeled dataset entry. 
Fig.~\ref{fig:soc_setups} compares these scenarios in terms of the Sum of Costs (SOC).
The following analysis is performed on the entire dataset of 12000 experiments. 

The average $SOC$ of the initial plans is $\SI{347.38}{\second}$. 
When executed in the real-time simulator, the average measured $SOC^{e}$ increases only slightly to $\SI{347.99}{\second}$, which means that the simulator does not introduce fundamental errors into the execution. 
The minor increase can be attributed to the overhead introduced by the \gls{ADG}, which coordinates up to 25 parallel threads corresponding to individual agents.

\begin{figure}[b]
  \centering
  \includegraphics[width=\columnwidth]{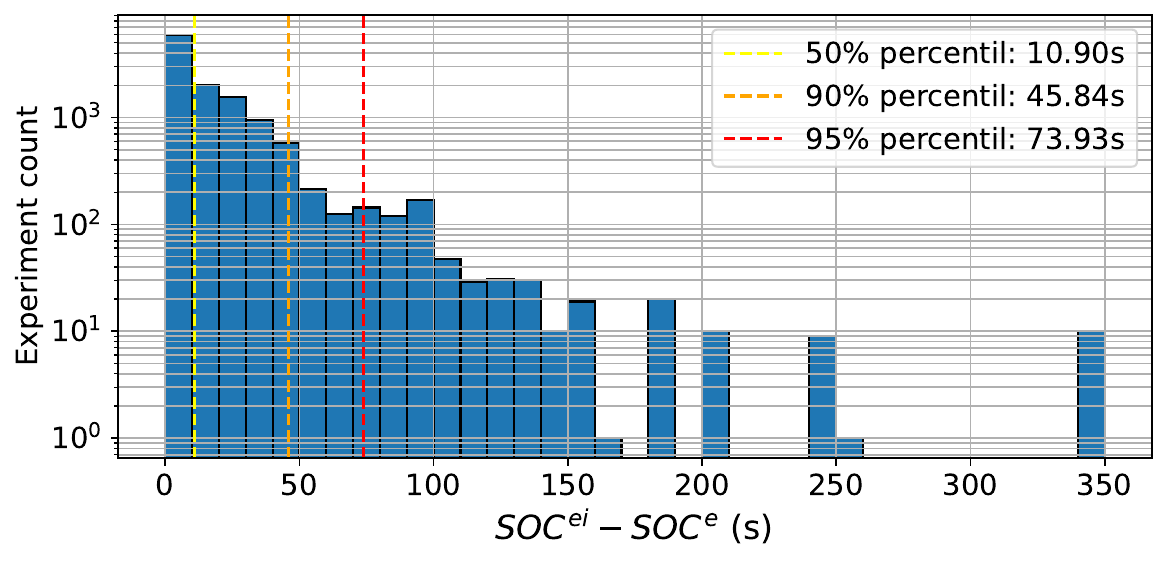}
  \caption{SOC increase caused by dynamic obstacle}
  \label{fig:soc_increase}
\end{figure}

Once the dynamic obstacle is introduced, the average $SOC^{ei}$ increases to $\SI{367.74}{\second}$, representing a $5.6\%$ increase compared to $SOC^{e}$. 
The absolute increase, computed as $SOC^{ei} - SOC^{e}$, is visualized in the histogram in Fig.~\ref{fig:soc_increase}. 
The plot shows that, for most runs, the dynamic obstacle has only a minor impact: in $50\%$ of all runs, the increase is below $\SI{10.90}{\second}$. 
However, in $5\%$ of the runs, the increase exceeds $\SI{73.93}{\second}$, corresponding to more than $20\%$ of the average $SOC^{e}$. 
The first objective of the \gls{RPP} is therefore to identify during execution these relatively rare but high-impact cases where the dynamic obstacle causes significant disturbance.
Some of the histogram bins in Fig.~\ref{fig:soc_increase} align closely with multiples of ten, which is expected since ten different replanning seeds are used per instance. 
This observation is consistent with the fact that the SOC increase is invariant to the randomly sampled replanning time $t$, although minor fluctuations may occur in the real-time simulator.

Let us now focus on the fourth boxplot in Fig.~\ref{fig:soc_setups}, which visualizes $SOC^{eir}_t$ - the total execution cost in the presence of a dynamic obstacle when replanning is triggered at the randomly sampled time $t$ and the resulting plan is executed until completion. 
The distribution is nearly identical to $SOC^{ei}$, leading to an important conclusion: although the solver is optimal, blindly 
replanning at an arbitrary time yields almost no benefit, even without considering the computational overhead. 
Moreover, $SOC^{eir}_t$ exhibits a higher number of outliers, indicating that in rare cases, replanning can be harmful. 
Two factors may explain this. 
First, the dynamic obstacle is unobservable. 
Thus, the new plan may be disrupted even more than the original one.
Second, some agents may be in the middle of executing an action at time $t$, causing the \gls{ADG} to delay the start of the new plan until those are completed, which can further increase the total cost.

Finally, the last boxplot in Fig.~\ref{fig:soc_setups} visualizes $SOC^{eirp}_t$, which also accounts for the computational overhead caused by replanning, as defined in Eq.~\ref{eq:overhead}. 
Its distribution is nearly identical to $SOC^{eir}_t$, which is a positive result: even though the solver is configured to return optimal solutions, it introduces only a negligible increase in the average execution cost for a single replanning. 
This observation, however, may not hold for significantly larger maps than those considered in this study, where each replanning may be more costly.

\begin{figure}[b]
  \centering
  \includegraphics[width=\columnwidth]{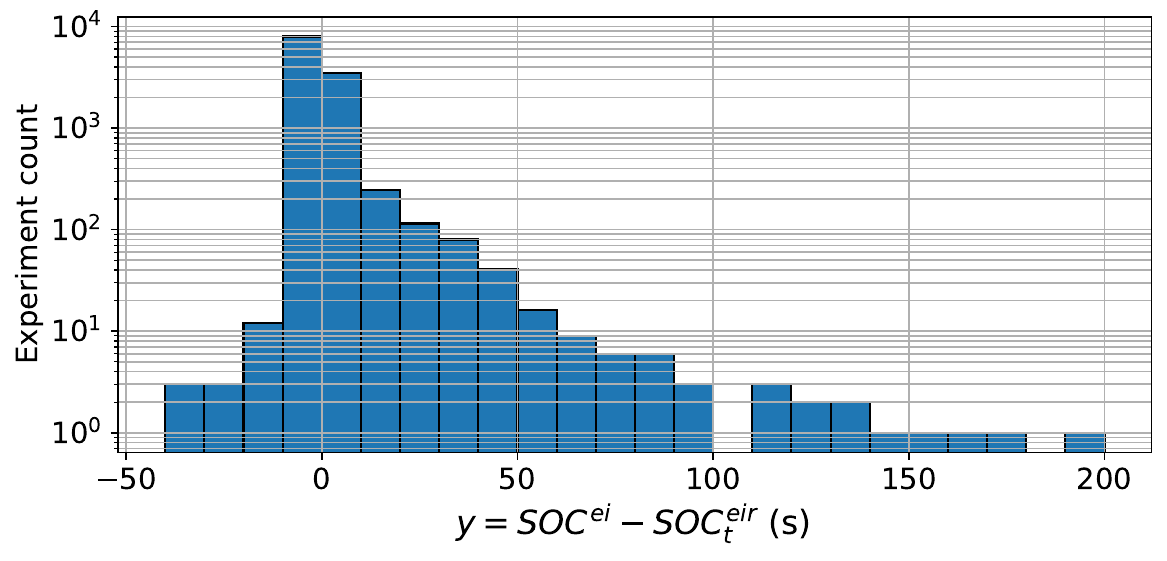}
  \caption{SOC savings achievable by random replanning}
  \label{fig:soc_savings}
\end{figure}

Since random replanning does not significantly improve the average case, the key question is whether it can provide benefits in extreme cases. 
Fig.~\ref{fig:soc_savings} addresses this question by showing the histogram of the regression target $y$, defined in 
Eq.~\ref{eq:target_2}, which represents the achieved savings. 
For a subset of $7.28\%$ of all instances, replanning yields savings of up to $\SI{200}{\second}$ - substantial given that the average $SOC^{e}=\SI{347.99}{\second}$. 
Thus, the \gls{RPP} aims not only to detect whether the disturbance has a significant impact on SOC, but also to determine whether replanning can mitigate it.

\subsection{Predicted SOC Savings}\label{sec:predicted_savings} 

\begin{figure}[h]
  \centering
  \includegraphics[width=\columnwidth]{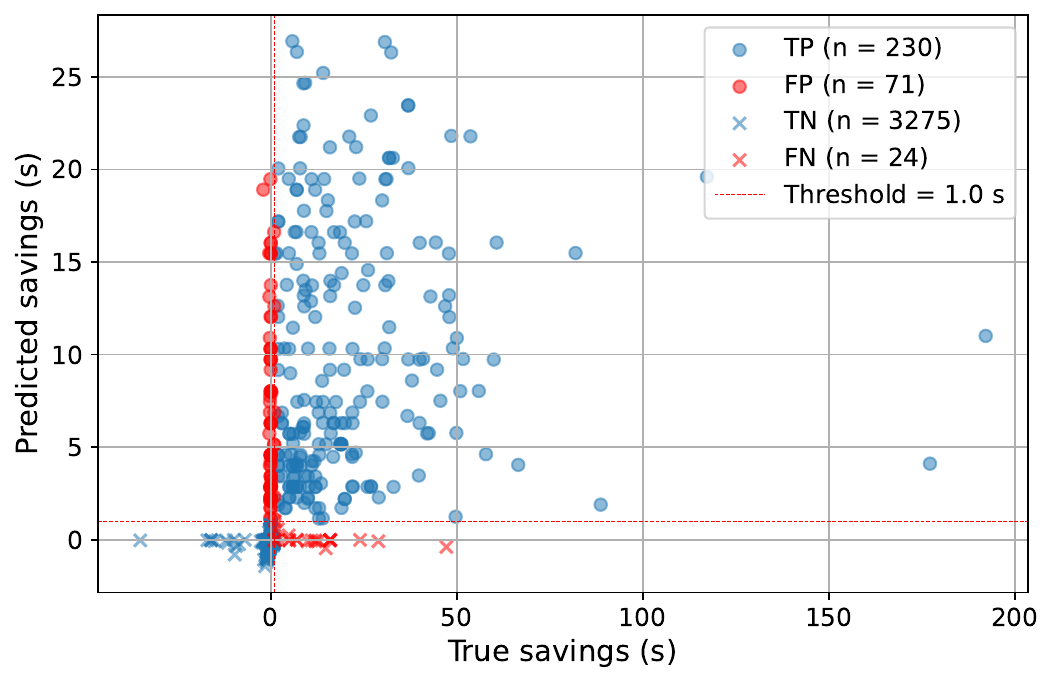}
  \caption{True vs. predicted SOC savings (test set)}
  \label{fig:soc_savings_predictions}
\end{figure}

In this section, we evaluate the prediction performance of the proposed approach. 
All experiments are conducted on the $\mathcal{D}_{test}$, which contains $3600$ samples, including $254$ positive cases (P), where replanning achieves a saving, and 3364 negative cases (N). 
The trained model achieves $\mathcal{L}_{\mathrm{MAE}}^{\mathrm{train}} = 1259.66$ on the training set and $\mathcal{L}_{\mathrm{MAE}}^{\mathrm{test}} = 1275.70$ on the test set. 
Fig.~\ref{fig:soc_savings_predictions} visualizes the predictions on the test set alongside their classification into positive and negative categories. 
The classification threshold is set to $\tau = \SI{1}{\second}$ to avoid treating small variations in runtime as meaningful savings.

The model achieves satisfactory sensitivity of $0.906$ and a specificity of $0.979$. 
Thus, selected positive and negative cases are very likely to be true positives and true negatives.
The precision is $0.764$, indicating a relatively high proportion of false positives. 
However, these false positives correspond to instances with true savings close to zero, meaning they trigger unnecessary replanning but do not have any harmful effect on final execution cost. 
The resulting F1 score is $0.829$.

\subsection{Realized SOC Savings}\label{sec:realized_savings}

\begin{figure}[t]
  \centering
  \begin{subfigure}{0.49\columnwidth}
    \centering
    \includegraphics[height=0.175\textheight]{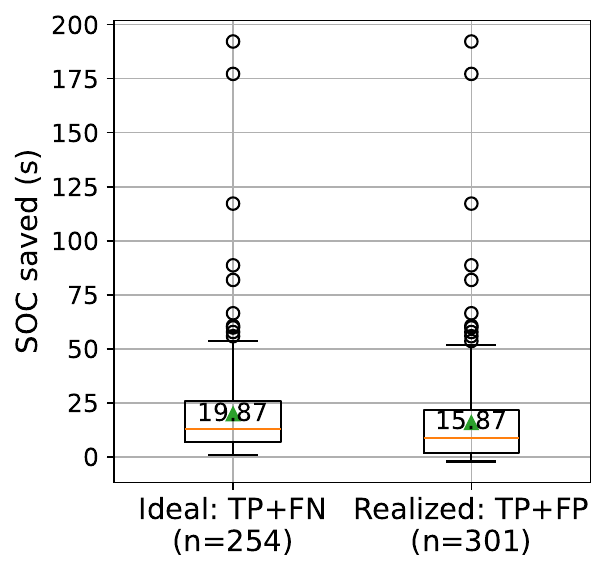}
    \caption{Absolute}
    \label{fig:soc_savings_absolute}
  \end{subfigure}\hfill
  \begin{subfigure}{0.49\columnwidth}
    \centering
    \includegraphics[height=0.175\textheight]{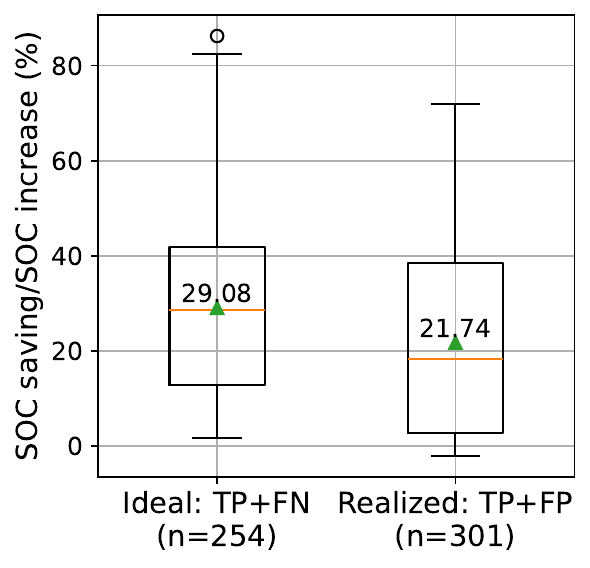}
    \caption{Relative}
    \label{fig:soc_savings_relative}
  \end{subfigure}
  \caption{Realized SOC savings per replanning}
  \label{fig:comparison}
\end{figure}

We now examine the \textit{realized SOC savings}, i.e., the actual reduction in $SOC^{ei}$ achieved when replanning is triggered according to the tuned model. 
These results are summarized in Fig.~\ref{fig:soc_savings_absolute}. 
When aggregated over $\mathcal{D}_{test}$, the realized replanning (TP + FP) achieved total savings of $\SI{4775.366}{\second}$ out of a potential $\SI{5047.336}{\second}$ (TP + FN), corresponding to a $94.6\%$ recovery rate.
In practice, our approach triggered replanning in $301$ cases (TP + FP), including $71$ false positives, where the realized saving is below the threshold $\tau = 1$ and can even be negative. 
In the ideal case of perfect prediction on the test set, only the $254$ positive cases (P = TP + FN) would be identified, yielding an average saving of $\SI{19.87}{\second}$ per experiment in which replanning is invoked. 
The average saving per realized replanning, which aggregates TP and FP, was $\SI{15.87}{\second}$.
We also report the SOC savings relative to the SOC increase in Fig.~\ref{fig:soc_savings_relative}. 
In the ideal case, replanning would recover on average only $29.08\%$ of the SOC increase across the $254$ experiments where a saving greater than the threshold $\tau$ is possible. 
With our approach, we achieve an average relative saving of $21.74\%$ over the $301$ experiments in which replanning was actually invoked.
In extreme cases, the realized replanning recovered more than $70\%$ of the SOC increase, corresponding to nearly $\SI{200}{\second}$. 
Conversely, the worst false positive resulted in a loss of only $\SI{2}{\second}$. 

\begin{figure}[b]
  \centering
  \begin{subfigure}{0.48\columnwidth}
    \centering
    \includegraphics[height=0.17\textheight]{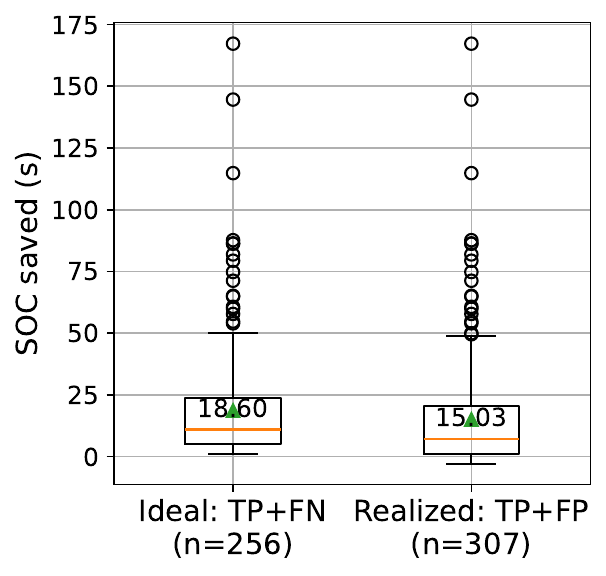}
    \caption{Absolute}
    \label{fig:soc_savings_absolute_w_solver}
  \end{subfigure}\hfill
  \begin{subfigure}{0.48\columnwidth}
    \centering
    \includegraphics[height=0.17\textheight]{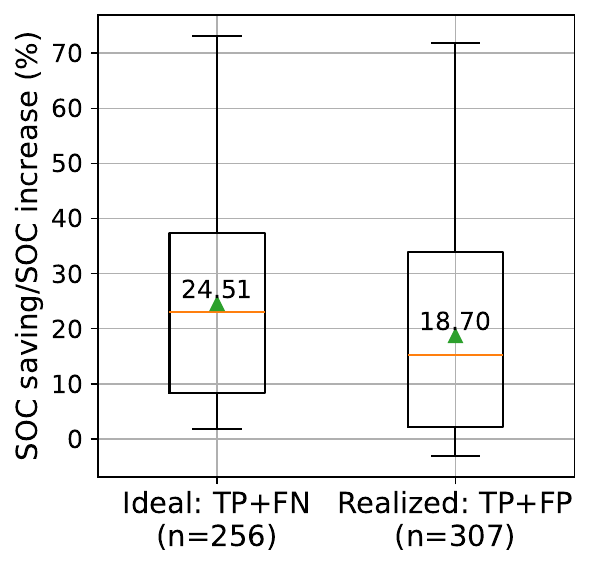}
    \caption{Relative}
    \label{fig:soc_savings_relative_w_solver}
  \end{subfigure}
  \caption{Realized SOC savings per replanning, including solver time}
  \label{fig:comparison_w_solver}
\end{figure}

Fig.~\ref{fig:comparison_w_solver} shows the realized SOC savings when the planning overhead is included, i.e., when $y$ 
from Eq.~\ref{eq:target_2} is redefined as $y = SOC^{ei} - SOC^{eirp}_t$. 
The results closely match those obtained when considering only the dynamic obstacle's impact (Fig.~\ref{fig:comparison}). 
This is a noteworthy finding, as the newly proposed features are designed to only quantify the disturbance caused by 
the dynamic obstacle, not to assess the computational difficulty of the replanning task or the associated overhead.

\subsection{Features importance}\label{sec:features_importance}

\begin{figure}[h]
  \centering
  \includegraphics[width=\columnwidth]{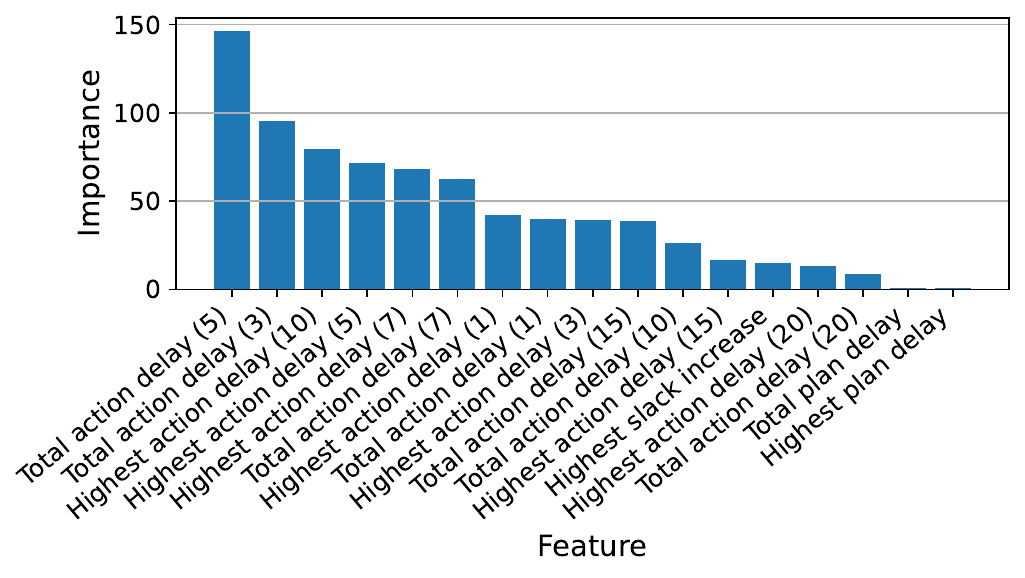}
  \caption{Features - permutation importance ($\geq 0.1$)}
  \label{fig:features_importance}
\end{figure}

In the final experiment, we analyze the importance of the individual features. 
Table~\ref{tab:features} lists the 18 proposed feature types, some of which are further parameterized with $n \in \{1, 3, 5, 7, 10, 15, 20\}$, 
resulting in a total of $42$ input features. 
Fig.~\ref{fig:features_importance} shows the permutation importance computed for the model trained on the test set. 
Permutation importance quantifies the increase in prediction error when the values of a given feature are randomly shuffled. 
For clarity, only features with an importance greater than $0.1$ are displayed.

The results indicate that only three feature types are strong predictors for triggering replanning: total action delay, highest action delay, and highest slack increase. 
Slack increase, which is relatively unimportant, was the only metric previously used to invoke replanning~\cite{zahradka2025holistica}.
Interestingly, static instance or plan properties, expected plan delays, and features describing the current execution progress, appear to be irrelevant.
The decision to replan is primarily driven by the magnitude of the already observed execution delays and their future propagation, reflected in slack increase.

\section{Conclusions \& Future Work}

We formulated the \emph{Replanning Prediction Problem (RPP)} and showed that a simple, planner-agnostic regression model can reliably decide 
when to replan during ADG-controlled MAPF execution with an unobservable dynamic obstacle. 
We introduced a bank of 18 feature types augmenting the ADG and, using a dataset of $12000$ experiments on four maps, demonstrated that random replanning offers little average benefit, whereas a minority of cases exhibit large recoverable losses. 
Our learned model effectively captures these high-impact cases: with a single replanning opportunity, it recovers $94.6\%$ of the available savings across the entire test dataset. 
In extreme cases, it recovers over $70\%$ of the execution cost per run, while false positives lead to negligible losses. 
Feature-importance analysis shows that decisions are driven primarily by observed action delays and their propagation (slack increase), whereas 
static instance or plan attributes contribute little. 
Notably, the method remains effective even when replanning computation overhead is included. 
Future work will extend the setting to sequential decisions with multiple replannings, larger maps where planning costs may dominate, and richer disturbance models (e.g., multiple or partially observable dynamic obstacles).

\bibliographystyle{IEEEtran}
\bibliography{main}

\end{document}